\documentclass[11pt,notoc,hyper]{article}
\usepackage{jcappub}
\usepackage{graphicx}
\usepackage{dcolumn}
\usepackage{bm,amsmath,amssymb} 
\usepackage{epsfig,multicol,bbm} 

\usepackage{latexsym}
\usepackage{epsfig}
\usepackage{amssymb}
\usepackage{graphicx}
\usepackage{amsmath}
\usepackage{graphicx}
\usepackage{amsfonts}
\usepackage{epsfig}
\usepackage{subfigure}

\title{Dynamics of three-form dark energy with dark matter couplings}
\author[a]{T.~Ngampitipan}
\author[a]{P.~Wongjun}
\affiliation[a]{Theoretical High-Energy Physics and Cosmology Group, Department of Physics,\\
Faculty of Science, Chulalongkorn University, Bangkok
10330, Thailand}
\emailAdd{pitbaa@gmail.com}
\arxivnumber{1108.0140}

\abstract{ Three-form field can give rise to viable cosmological scenarios of dark energy with potentially observable signatures distinct from standard single scalar field models. In this study, the background dynamics of three-form cosmology are investigated. Our analysis suggests that the potential of three-form field should be runaway. We also investigate the possibility to solve the coincidence problem by using the coupling of three-form field to dark matter. Four types of coupling form are examined and we find that the coupling of the form $Q = \sqrt{6}\Gamma (\dot{X} + 3HX)/\kappa$ provides the possibility to solve the coincidence problem.}

\begin{document}
\maketitle

\section{Introduction}
According to cosmological observations, today our universe has the accelerated expansion \cite{dark energy1, dark energy2}. What produces that acceleration is the so-called \textbf{dark energy}. Moreover, there is \textbf{dark matter} inferred from the existence of gravitational effects on visible matter and background radiation, but is undetectable by emitted electromagnetic radiation \cite{dark matter1, dark matter2}. Recent observations imply that today 72\% of all the energy densities of the universe comes from dark energy, while dark matter contributes to 23\%. The remaining 5\% is well-known matter and radiation \cite{DE DM1, DE DM2}.

There are many models of dark energy. The most useful and simple model is cosmological constant. The results of dark energy from cosmological constant can properly fit the observations, but it encounters cosmological constant problem and coincidence problem. Since the energy densities of cosmological constant and dark matter are significantly different throughout the history of the universe, the problem is posed why their energy densities are of the same order at the present time. That problem is referred to the coincidence problem. The dark energy models with varying energy density are proposed in order to solve the problems of cosmological constant. There are many models explaining dark energy and most of them are based on the use of scalar field \cite{quintessence1, quintessence2}. However, an uncoupled scalar field cannot account for the coincidence problem \cite{uncoupled scalar field}. One of the possibilities to handle the coincidence problem is that the scalar field should couple to dark matter \cite{int1, Wetterich} and the coupling can lead to an accelerated scaling attractor solution \cite{Maartens}.

However, the experimental evidences of the fundamental scalar particle have not been discovered. Moreover, at a more fundamental level there is no reason to exclude the possibility of some other higher form field to be the dark energy. These higher form fields can exist in some theories such as the string theory. The presence of them does not necessarily violate the cosmological principle. For these reasons an effort has been made in using a vector field, a one-form field, to play a role of dark energy \cite{VV, CA, CG, MN}. However, most of the vector field models encounter instabilities \cite{BH}. Generalizations to higher form fields have also been proposed \cite{Gm, nform}. Two-form field models also have the same problem as vector field models. On the other hand, three-form field models have no such problem: they are stable \cite{nform}. Therefore, it is of interest to consider a three-form field as a candidate for dark energy. Recently, dark energy from three-form field has been studied and it is found that three-form field can provide the accelerating universe \cite{Koivisto}. However, the coincidence problem cannot be solved since there are no stable fixed points at which energy densities of dark matter and dark energy are comparable.

According to interacting scalar models, there are three classes of coupling form, $Q_{I} \propto \dot{\phi} \rho_{c}$, $Q_{II} \propto H(\alpha_\phi\rho_{\phi} + \alpha_c\rho_{c})$, and $Q_{III} \propto (\alpha_\phi \rho_{\phi} + \alpha_c\rho_{c})$ where subscript $\phi$ and $c$ denote the contribution for the scalar field and cold dark matter respectively. For the first type, it cannot provide the standard evolution of the universe while with the same parameters the coincidence problem is solved \cite{shinji}. For the second and third types with exponential potential and $\alpha_\phi = 0$ case, the coincidence problem cannot be solved since there are no stable fixed points at which $\Omega_\phi/\Omega_c \sim O(1)$ \cite{Maartens}. However, in the case of $\alpha_\phi \neq 0$ and $\omega_\phi = \text{constant}$, there are the fixed points which provide the possibility to solve the coincidence problem. Nevertheless, they also yield the negative energy density of the dark energy \cite{int2 negativeE, int23 negativeE}. From many investigations of the interacting scalar field dark energy, it suggests that it is not easy to solve the coincidence problem with the proper behavior of the evolution. Therefore, it is worthwhile to investigate whether the interacting three-form field can provide the possibility to solve the coincidence with the proper behavior of the evolution and this is the aim of our paper.

It is important to note that the coupled quintessence with growing matter, a matter consisting of particles with an increasing mass, provides one of the possible solutions to solve the coincidence problem \cite{growing matter1}. The candidate of the growing matter is suggested as neutrinos and then the observed dark energy density and the equation of state parameter are determined by the neutrinos mass. Significant results of this model are investigated and compared with the observational data \cite{growing matter2}.

We review dark energy from non-interacting three-form field in section \ref{three-form}. Our analysis suggests that the potential of the three-form field should be runaway. This suggestion is obtained by ensuring the well-behaved evolution of the universe. Analogous to scalar field dark energy, three types of simple coupling form of the three-form field with dark matter are examined for various runaway potentials in section \ref{coupled}. In order to provide the consistent perturbations in the models, three types of the coupling form are introduced with the ability to express in the covariant form. It is found that the coincidence problem cannot be solved. To deal with the coincidence problem, we introduce another covariant coupling form in section \ref{extension}. It is found that this coupling form can provide the ability to solve the coincidence problem with the well-behaved evolution. The results of our investigation are summarized and discussed in section \ref{Conclusions}.

\section{Dark Energy from Three-form Field}\label{three-form}

Due to the homogeneous and isotropic universe in the large scale,  we consider the flat FRW universe with the line element
\begin{equation}
ds^{2}=-dt^{2}+a^{2}(t)d\textbf{x}^{2}.\label{frw}
\end{equation}
We assume that the universe consists of matter (both baryon and cold dark matter), radiation, and dark energy. We describe each of these species as a fluid. In this paper, the fluid is assumed to be perfect, the fluid with no viscosity and momentum density. For the perfect fluid, the energy momentum tensor takes the form
\begin{equation}
T^{\mu}_{\;\ \nu}=Diag(-\rho, p, p, p),
\end{equation}
where $\rho$ is the energy density of the fluid and $p$ the pressure. The Einstein field equation gives two equations
\begin{eqnarray}
H^{2}\equiv\Big(\frac{\dot{a}}{a}\Big)^{2}=\frac{\kappa^2}{3}\sum_{i}\rho_i,\label{G00}\\
\dot{H}=-\frac{\kappa^2}{2}\sum_{i}(\rho_i+ p_i),\label{Gii}
\end{eqnarray}
where $H$ is the Hubble parameter and $\kappa^2 = 8 \pi G$. The conservation of the energy momentum tensor for each species can be expressed as
\begin{equation}
\dot{\rho_i}+3H(\rho_i + p_i)=0.\label{con}
\end{equation}
Using (\ref{G00}) and (\ref{Gii}), we obtain the acceleration equation
\begin{equation}
\frac{\ddot{a}}{a}=-\frac{\kappa^2}{6}\sum_{i}(\rho_i+3p_i)=-\frac{\kappa^2}{6}\sum_{i}\rho_i(1+3\omega_i),
\end{equation}
where $\omega_i$ is the equation of state parameter for each species. Therefore, the accelerating universe due to each species requires that
\begin{equation}
\omega_i < -1/3,\label{acc}
\end{equation}
where, $\omega_{\text{matter}} = 0$ and $\omega_{\text{radiation}} = 1/3$. Thus both matter and radiation cannot yield the accelerating universe that we observe nowadays. In order to explain the current accelerating universe, a new constituent of the universe is introduced namely \textit{Dark Energy}. There are many models of dark energy which use scalar field such as quintessence, tachyon, phantom \cite{reviewDE} and chameleon \cite{chameleon_Mota, chameleon}. Recently, three-form field is introduced as a candidate of dark energy models \cite{Koivisto}. We will review the dynamical properties of three-form field dark energy model in this section. These properties are also compared with the results of the quintessence model. We comment on the drawbacks and advantages of the model from the point of view of observational compatibility.

We are interested in a theory of minimal coupling to gravity. The action of a three-form field $A_{\mu\nu\rho}$ is then
\begin{equation}
S_{A}=-\int d^4x\sqrt{-g}\Big(\frac{1}{2\kappa^{2}}R-\frac{1}{48}F^{2}-V(A^{2})\Big),\label{1}
\end{equation}
where $R$ is the Ricci scalar  and $V(A^{2})$ is the potential of the field and $A^{2}\equiv A^{\mu\nu\rho}A_{\mu\nu\rho}$. $F_{\mu\nu\rho\sigma}$ is the field strength tensor of the three-form field defined as
\begin{equation}
F_{\mu\nu\rho\sigma}\equiv4\nabla_{[\mu}A_{\nu\rho\sigma]},
\end{equation}
where the square bracket means antisymmetrization. Varying the action (\ref{1}) with respect to $A^{\mu\nu\rho}$, we get the equation of motion
\begin{equation}
\nabla^{\alpha}F_{\alpha\mu\nu\rho}=12V'(A^{2})A_{\mu\nu\rho},\label{eom}
\end{equation}
where $V'(A^{2}) \equiv \text{d}V/\text{d}\left(A^{2}\right)$. The Einstein field equation is obtained by varying the action (\ref{1}) with respect to $g^{\mu\nu}$,
\begin{equation}
G_{\mu\nu}\equiv R_{\mu\nu}-\frac{1}{2}R g_{\mu\nu}=\kappa^2 T_{\mu\nu},\label{ein}
\end{equation}
where $G_{\mu\nu}$ is the Einstein tensor, $R_{\mu\nu}$ is the Ricci tensor, and $T_{\mu\nu}$ is the energy momentum tensor given by
\begin{equation}
T_{\mu\nu}=\frac{1}{6}F_{\mu\alpha\beta\gamma}F_{\nu} ^{\alpha\beta\gamma}+6V'(A^{2})A_{\mu\alpha\beta}A_{\nu} ^{\alpha\beta}-g_{\mu\nu}\Big(\frac{1}{48}F_{\alpha\beta\gamma\rho}F^{\alpha\beta\gamma\rho}+V(A^{2})\Big).
\end{equation}
For the FRW metric (\ref{frw}), where a three-form field depends only on time, a zero component of three-form field is nondynamical. From its equation of motion (\ref{eom}), it has algebraic constraint $12V'(A^{2})A_{0\mu\nu} = 0$. Therefore we focus on the space-like components only. In this spacetime with the homogeneity and isotropy assumed, the space-like components take the form
\begin{equation}
A_{ijk}=a^{3}(t)\epsilon_{ijk}X(t),
\end{equation}
where $X$ is a comoving field. The field $X$ is related to the field $A^{\mu\nu\rho}$ via $A^{2}=6X^{2}$. From (\ref{eom}) the equation of motion of the field $X$ is then
\begin{equation}
\ddot{X}=-3H\dot{X}-V,_{X}-3\dot{H}X,\label{3formeom}
\end{equation}
where overdot denotes derivative with respect to the time $t$ and $V,_{X}\equiv \text{d}V/\text{d}X$. The energy density and pressure of the three-form field are given by
\begin{eqnarray}
\rho_{X}&=&-T^{0}_{0}=\frac{1}{2}(\dot{X}+3HX)^{2}+V(X)\label{rhoX}\\
p_{X}&=&T^{i}_{i}=-\frac{1}{2}(\dot{X}+3HX)^{2}-V(X)+V,_{X}X.\label{pX}
\end{eqnarray}
The equation of state parameter of the three-form field is
\begin{equation}
\omega_{X}=\frac{p_{X}}{\rho_{X}}=-1+\frac{V,_{X}X}{\rho_{X}}\label{eosp}.
\end{equation}
Conveniently, we will consider the dimensionless variables for characterizing the dynamics of the three-form field. From equation (\ref{Gii}), (\ref{con}) and (\ref{3formeom}), the dynamical equations are
\begin{eqnarray}
x' &=& 3(\sqrt{\frac{2}{3}}y - x),\label{dynamics1}\\
y' &=& -\frac{3}{2}\lambda(x)z^{2}(x y-\sqrt{\frac{2}{3}})+\frac{3}{2}(w^{2}+\frac{4}{3}u^{2}+v^{2})y,\label{dynamics2}\\
v' &=& -\frac{3}{2}v[1+\lambda(x)xz^{2}-w^{2}-\frac{4}{3}u^{2}-v^{2}],\label{dynamics3}\\
w' &=& -\frac{3}{2}w[1+\lambda(x)xz^{2}-w^{2}-\frac{4}{3}u^{2}-v^{2}],\label{dynamics4}\\
u' &=& -\frac{3}{2}u[\frac{4}{3}+\lambda(x)xz^{2}-w^{2}-\frac{4}{3}u^{2}-v^{2}],\label{dynamics5}
\end{eqnarray}
where the dimensionless variables are defined as
\begin{eqnarray}
x\equiv\kappa X,~~~~y\equiv\frac{\kappa}{\sqrt{6}}(X'+3X),~~~~z^{2}\equiv\frac{\kappa^{2}V}{3H^{2}},~~~~w^{2}\equiv\frac{\kappa^{2}\rho_{b}}{3H^{2}},\nonumber\\ u^{2}\equiv\frac{\kappa^{2}\rho_{r}}{3H^{2}}, ~~~~v^{2}\equiv\frac{\kappa^{2}\rho_{c}}{3H^{2}},~~~~\lambda(x)\equiv-\frac{1}{\kappa}\frac{V,_{X}}{V},
\end{eqnarray}
where $\rho_{b}$ is energy density of baryon, $\rho_{r}$ is energy density of radiation, and $\rho_{c}$ is energy density of cold dark matter. The prime denotes the derivative with respect to $N\equiv\ln a$, the $e$-folding number. Equation (\ref{G00}) is not the dynamical equation. It is the constraint equation expressed as
\begin{eqnarray}
y^2 + z^2 + u^2 + v^2 + w^2 = 1.
\end{eqnarray}
We can find the fixed points of the system by setting $x'$, $y'$, $v'$, $w'$, and $u'$ equal to zero. We summarize all interesting fixed points in Table \ref{fix pt 3form}.
\begin{table}
\centering
\begin{tabular}{|c|c|c|c|c|c|c|c|c|}
  \hline
  & &  & &  &  & & & \\
  Name & $x$ & $y$ & $z$ &$v$ & $w$ & $u$ &$\omega_{\text{tot}}$ & Existence \\\hline
  & &  & &  &  & &  &\\
  (a) & 0 & 0 & 0 &$\pm1$ & 0 & 0 & 0 & all $\lambda$  \\\hline
  & &  & &  &  & &  &\\
  (b1) & $\sqrt{\frac{2}{3}}$ & 1 & 0&0 & 0 & 0 &-1 & all $\lambda$  \\\hline
  & &  & &  &  & &  &\\
  (b2) & $-\sqrt{\frac{2}{3}}$ & -1 & 0&0 & 0 & 0 &-1 & all $\lambda$  \\\hline
  & &  & &  &  & &  &\\
  (c) & $\sqrt{\frac{2}{3}}y_*$ & $0 \leq y_* \leq 1$ & $\sqrt{1-y_*^2}$& 0 & 0 & 0 & -1 & $\lambda=0$ \\\hline
\end{tabular}
\caption{Fixed points in the three-form field dark energy model.}\label{fix pt 3form}
\end{table}

Now we have all equations for characterizing the dynamics of the three-form field. The first interesting point is the equation of state parameter of three-form field in equation (\ref{eosp}). It satisfies the condition of accelerating universe in (\ref{acc}). Moreover, it can mimic the cosmological constant behavior in the case of vanishing of the potential slope. The equation of state parameter can also be adjusted as the phantom phase universe ($\omega < -1$) in order to satisfy the recent observational data, $\omega = -1.1 \pm 0.14$ \cite{wmap7}. Actually, it can take any real value.

Next, we will consider the dynamical equations analytically. From the fact that there are matter- and radiation-dominated periods, this yields the result that the three-form field dark energy needs to sub-dominate in those periods. This argument will suggest us what the shape of the potential is. By considering the matter-dominated period (setting $w=0,y=0,u=0, z=0$), one can get the fixed point giving that period from the dynamical equations (\ref{dynamics2}) - (\ref{dynamics5}). However, from equation (\ref{dynamics1}), the value of $x$ is not constrained. At $y=0$, it grows exponentially when the time is back to the past (N is more negative). This means that $x$ must be a huge value at that time. At the matter dominated-period, $z$ needs to vanish. This leads to the vanishing of the potential at a large value of $X$. This suggests us that the potential of three-form field should be runaway. However, if we look at the simple runaway potential such as an exponential potential, it will not vanish at the finite $X$. We will discuss this issue by taking $w=0, y=0, u=0$ and $ z=\varepsilon$ where $\varepsilon$ is a small number. The dynamical equations (\ref{dynamics4}), and (\ref{dynamics5}) vanish automatically. The dynamical equations (\ref{dynamics2}) and (\ref{dynamics3}) can be written as
\begin{eqnarray}
y' &=& \sqrt{\frac{3}{2}} \lambda(x)\varepsilon^{2},\label{analys1}\\
v' &=& -\frac{3}{2}\sqrt{1-\varepsilon^{2}}\Big( 1 + x\lambda(x)\Big)\varepsilon^{2}.\label{analys2}
\end{eqnarray}
In order to ensure that there is the matter-dominated period, $y'$ must be positive and $v'$ must be negative at the time after matter-dominated period (If the time is back to the past, it is the time before matter-dominated period). For $x \gg 1$, one obtains the constraint
\begin{eqnarray}
\lambda \gtrsim 0.
\end{eqnarray}
This constraint is valid even when we consider radiation-dominated period. Again, the constraint of $\lambda$ suggests us that the three-form potential should be runaway.

We note that the runaway potential we mentioned means that the value of the potential at the large $X$ approaches zero $\displaystyle \lim_{X\rightarrow\infty} V(X) \rightarrow 0$ and the slope of the potential is always negative $V_{,X} < 0$. The runaway potential will guarantee the well-behaved evolution of the background. However, some kinds of non-runaway potential will also provide the well-behaved evolution. For example, the power law potential, $V= V_0 x^{n}$, with $n=1$ also provides the well-behaved evolution while the Landau-Ginzburg potential, $V= V_0 (x^{2} - x_0^2)^2$, cannot provides such well-behaved evolution. We will consider this issue in detail. Generally, we can set the value of $z^2$ and $y^2$ to be a small number $\varepsilon^{2}_z$ and $\varepsilon^{2}_y$ respectively. Therefore equations (\ref{analys1}) and (\ref{analys2}) can be generalized as
\begin{eqnarray}
y' &=& \frac{3}{2}\Big( \lambda(x)\varepsilon^{2}_z\sqrt{\frac{2}{3}} + \varepsilon_y \Big),\label{analysgen1}\\
v' &=& -\frac{3}{2}\sqrt{1-\varepsilon_z^{2}-\varepsilon_y^{2}}\Big( 1 +\frac{\varepsilon_y^{2}}{\varepsilon_z^{2}}+ x\lambda(x)\Big)\varepsilon_z^{2}.\label{analysgen2}
\end{eqnarray}

For the power law potential, $V= V_0 x^{n}$, it ensures that the well-behaved evolution will be obtained if $n \leq 1$ where $\lambda = -n/x$. Actually $n$ can take the value larger than $1$ depending on the value of $\varepsilon_y^{2}/\varepsilon_z^{2}$. We can see that the non-runaway potential, $ 0 \leq n \leq 1$, can also provide well-behaved evolution.

For Landau-Ginzburg potential, $V= V_0 (x^{2} - x_0^2)^2$, with $\lambda = -4x/(x^{2} - x_{0}^{2})$, one cannot get the well-behaved evolution at the large $X$.  Nevertheless, one can argue that the value of $X$ will be a small value, ($x \ll x_0$), if $\sqrt{2/3}y$ always grater than $x$ inferred from equation (\ref{dynamics1}). However, it is difficult to achieve this situation since $y$ decreases faster than $x$ when the time is back to the past during three-form dominated period. We also note that we check this analysis with the numerical calculation and found that it agrees with this qualitative analysis.

At the matter- and radiation-dominated periods, we see that the three-form field $X$ is a large value. This leads to a huge negative value of $\dot{X}$ as $y \propto \dot{X} + 3 H X = 0$. Instead a scalar field rolls down from the small value like quintessence field, the three-form field rapidly climbs up the potential from the large value of field to the small one and reaches the fixed point at which dark energy dominates at the present time. One of the most interesting things in the three-form dark energy model is the effective potential. From the equation of motion of three-form field (\ref{3formeom}) and equation (\ref{Gii}), the effective potential can be defined through the relation
\begin{equation}
V_{eff},_{X} = (1-\frac{3}{2}x^2)V,_{X}-\frac{3}{2}\kappa x \sum_{i} \gamma_i \rho_i.\label{Veff}
\end{equation}
At the dark energy dominated period, we can neglect the last term which is the contributions from matter and radiation. This means that the fixed point at the extremum of the potential is $x=\sqrt{2/3}$ which is the same as we consider from dynamical equation (\ref{dynamics1}). This is a mechanism by which the potential parameter controls the dynamics of three-form field. It is similar to the analysis in reference \cite{Koivisto}. Note that in \cite{Koivisto}, for the potential $V= (X^2 - C^2)^2$, the dynamics of three-form field is controlled by $C$ which is less or more than $\sqrt{2/3}$.

Now, let us analyze the stability of the fixed points. The fixed point (a) corresponds to the matter-dominated solution and gives $\omega_{\text{tot}} = 0$. The eigenvalues are $(-3, 3/2, 3)$; therefore, it is not stable. Note that we do not include radiation and baryon in this investigation. However, if we add all constituents in our investigation, there will be the unstable fixed points in which the added constituents dominate. Properties of these fixed points are similar to those of fixed point (a). The eigenvalues of these fixed points can be generalized as $[\;-3, 3(1 + \omega_i)/2, 3(1 + \omega_i)\;]$, where $\omega_i$ is the equation of state parameter of the dominated constituent $i$.

The fixed points (b) correspond to the three-form field dominated solution and give $\omega_{\text{tot}} = -1$. Therefore, three-form field behaves as dark energy. They give the eigenvalues $(-3, 0, -3/2)$. Since one of the eigenvalues is zero, we can say nothing about the stability of the fixed points from the linear analysis. We have to use the second order perturbations and specify the potential to analyze the stability of these fixed points. However, we will use the numerical method to analyze the property of the fixed points. In the numerical method, we check whether the small deviations of the dynamical variables are convergent and also check whether their evolution provides the standard evolution of the universe. Note that the standard evolution of the universe we use in this paper means that there are the radiation- matter- and dark energy-dominated periods respectively. For the exponential potential, $V = V_0 \exp(-\eta x)$, we find that the fixed point (b1) is stable for $\eta > 0$ and the fixed point (b2) is stable for $\eta < 0$. For inverse power law potential, $V = V_0 x^{-\eta}$ and Gaussian potential $V = V_0 \exp(-\eta x^{2})$ both of the fixed points (b) are stable for $\eta \gtrsim -1$ and $\eta > 0$ respectively. These properties of the fixed points (b) are consistent to our analytic investigation that the potential should be runaway and it is compatible with the second order perturbation analysis in \cite{Koivisto}.

The fixed points (c) provides the cosmological constant behavior since it corresponds to $\lambda = 0$ for runaway potential. This is the special case of the runaway potential since it yields the constant energy density of dark energy due to $V = V_0$. We note that it may not be completely correct to mention that it is runaway potential when it is constant. This constant potential leads to the stable fixed point with $y^2 + z^2 = \Omega_X = 1$, $\omega_{\text{tot}} = -1$ and $\omega_{\text{X}} = \text{constant} = -1$. This fixed point corresponds to the extremum fixed points in \cite{Koivisto}. We avoid the call this fixed point as extremum fixed point since it is constant potential. We also note that there is the extremum fixed point for Gaussian potential with $x = 0, y = 0, z = 1$ and $\lambda \neq 0$. This fixed point will be stable if $\eta < 0$. However, it leads to the dark energy dominated period in the past.

Cosmological constant model can be promoted as a subclass of three-form dark energy model. This is a useful feature of dark energy from three-form field. However, it cannot solve the coincidence problem which state that ``why dark energy is comparable to matter nowadays". This is due to the fact that there are no the stable fixed points at which $\Omega_X /\Omega_c \simeq 7/3$ where $\Omega_X$ and $\Omega_c$ are the energy density parameter for dark energy and dark matter respectively. Next section, we will investigate an probability for solving this problem by introducing the interactions between dark energy and dark matter.

\section{Coupling Three-form Field with Dark Matter}\label{coupled}

According to observations, today the dark matter density is close in value to the dark energy density. This leads to the so-called coincidence problem because their evolutions are considerably different throughout the universe history. Is it coincidence that their energy densities are of the same order? There have been many models proposed to explain the coincidence problem and it has been found that an uncoupled three-form field cannot solve the coincidence problem. If dark matter is capable of decaying into dark energy, the explanation of the similarity of their energy densities may be made. This introduces the coupling between three-form field dark energy and dark matter. In order to alleviate the coincidence problem we expect this coupling to lead to an accelerated scaling attractor solution
\begin{equation}
\frac{\Omega_{\text{dark energy}}}{\Omega_{\text{dark matter}}} = \textit{O}(1) ~~\text{and}~~ \ddot{a} > 0.
\end{equation}
The existence of the coupling can be represented by the modified continuity equations
\begin{eqnarray}
\dot{\rho}_c &=& -3H\rho_c - Q,\label{rhodotc}\\
\dot{\rho}_X &=& -3H(\rho_X + p_X) +Q\label{rhodotX},
\end{eqnarray}
where $\rho_{c}$ stands for energy density of cold dark matter and $Q$ is the energy transfer between dark energy and dark matter
\begin{center}
$Q>0\Rightarrow$ dark matter $\rightarrow$ dark energy\\
$Q<0\Rightarrow$ dark energy $\rightarrow$ dark matter,
\end{center}
while the background baryons and radiation still satisfy
\begin{eqnarray}
\dot{\rho_{b}} &=& -3H\rho_{b},\\
\dot{\rho_{r}} &=& -3H(\rho_{r}+p_{r}),
\end{eqnarray}
where $b$ stands for baryon and $r$ radiation. The explicit form of the Einstein equations (\ref{G00}) and (\ref{Gii}) including baryons and radiation become
\begin{eqnarray}
H^{2} &=& \frac{\kappa^{2}}{3}\left[\frac{1}{2}(\dot{X} + 3HX)^{2} + V(X) + \rho_{b} + \rho_{r} + \rho_{c}\right]\\
\dot{H} &=& -\frac{\kappa^{2}}{2}(V,_{X}X + \rho_{b} + \rho_{r} + p_{r} + \rho_{c}).
\end{eqnarray}
From equation (\ref{rhodotX}), using (\ref{rhoX}) and (\ref{pX}) we obtain
\begin{equation}
\ddot{X}+3H\dot{X}+3\dot{H}X+V,_{X}=\frac{Q}{\dot{X}+3HX}.\label{eomX}
\end{equation}
The interaction term we added in equations (\ref{rhodotc}) and (\ref{rhodotX}) will modify the dimensionless variable equations (\ref{dynamics2}) and (\ref{dynamics3}) as
\begin{eqnarray}
y' &=& \overline{\gamma}-\frac{3}{2}\lambda(x)z^{2}(xy-\sqrt{\frac{2}{3}})+\frac{3}{2}(w^{2}+\frac{4}{3}u^{2}+v^{2})y,\label{auto2}\\
v' &=& -\frac{\overline{\gamma}y}{v}-\frac{3}{2}v[1+\lambda(x)xz^{2}-w^{2}-\frac{4}{3}u^{2}-v^{2}],\label{auto3}
\end{eqnarray}
where
\begin{eqnarray}
\overline{\gamma}=\frac{\kappa Q}{\sqrt{6}(\dot{X}+3HX)H^{2}}.
\end{eqnarray}
Coupling forms in which we use to consider in this paper are phenomenological. The coupling forms are introduced by taking into account the fact that it should be expressed in the covariant form due to the ability to compute the cosmological perturbations. According to the quintessence model, there are three simple coupling forms. The first coupling is $Q \propto \dot{\phi}\rho_{c}$, where $\phi$ is the quintessence field. This coupling form is motivated from the scalar-tensor theory \cite{int1}. Unfortunately, by considering the general form of Lagrangian of scalar field with this coupling form, there are no fixed points giving the matter- or radiation-dominated periods if dark energy dominates at the present time \cite{shinji}. Analogous to this coupling form of the scalar field and its simple kinetic term, we introduce the coupling form of the three-form field by replacing $\dot{\phi}$ with $\dot{X} + 3H X$. Note that there is no theoretical motivation like the quintessence models.

The second class of the coupling is $Q \propto H(\alpha_\phi \rho_{\phi} + \alpha_c \rho_{c})$, where $\rho_{\phi}$ is energy density of the scalar field and $\alpha_{\phi}$ and $\alpha_{c}$ are dimensionless constants. The motivation of this coupling form does not come from the fundamental theory but mathematical calculation in which the energy density ratio of dark matter to dark energy is constant at a stable fixed point \cite{int2}. In this coupling form, there is no the problem like the previous one but it gives a negative energy density of dark energy at matter- and radiation-dominated period \cite{int2 negativeE}. By considering $\alpha_\phi = 0$ case with exponential potential, the coincidence problem cannot be solved since there are no stable fixed points $\Omega_\phi/\Omega_c \sim O(1)$ \cite{Maartens}. Note that the transfer rate of this coupling form is determined by the expansion rate of the universe rather than by the local quantities associated with particle interactions. Therefore, it suggests that it is not easy to describe this coupling form in terms of physical interaction of particles.

The third coupling form is introduced in order to obtain the physical motivation associated with particle interactions. The similar form of this coupling has been use in reheating, decay of curvaton and dark matter into radiation \cite{Maartens}. This coupling takes the form $ Q \propto (\Gamma_\phi\rho_{\phi} + \Gamma_c \rho_{c})$, where $\Gamma_i$ is a constant. This coupling form does not depend on the expansion rate of the universe. $\Gamma$ acts as a local transfer rate which is similar to the decay rate of the particle interaction in collision approach. This form is quite complicated because one cannot eliminate the Hubble parameter from the dynamical equations of the dimensionless variables. Similar to second coupling form, it encounters the negative energy density in full parameters consideration ($\Gamma_\phi \neq 0$ and $\Gamma_c \neq 0$) while it cannot solve the coincidence problem in the case of $\Gamma_\phi = 0$ \cite{int23 negativeE}.

It is worthwhile to investigate whether interacting three-form field can yield the solution of the coincidence problem. In order to study the three-form interaction with dark matter phenomenologically, we introduce the covariant forms of the coupling for three-form field as:
\begin{center}
\begin{tabular}{cl}
(I) &   $Q^\mu = \sqrt{2/3}\kappa\beta T^\tau_{\tau(c)} \left(1/24a^3\right)\epsilon^{\nu\rho\sigma\gamma}F_{\nu\rho\sigma\gamma} u^{\mu}$,  \\
(II) &    $Q^\mu =\alpha H T^\tau_{\tau(c)} u^{\mu}$, \\
(III) &  $Q^\mu = \Gamma T^\tau_{\tau(c)} u^{\mu}$,\\
\end{tabular} \end{center}
where $\alpha$, $\beta$ and $\Gamma$ are constant, $u^{\mu}$ is a 4-velocity, $T^\tau_{\tau(c)} = -\rho_c$ is the trace of the energy-momentum tensor for cold dark matter. $Q^{\mu}$ is the covariant form of energy-momentum transfer corresponding to the covariant form of the conservation equation \cite{covariant EM tensor}
\begin{equation}
\nabla_{\nu}T^{\mu\nu}_{i} = Q^{\mu}_{i},\label{covariant conservation}
\end{equation}
with
\begin{equation}
\sum_{i}Q^{\mu}_{i} = 0,
\end{equation}
where $i$ is a component of the universe. The zero component of the covariant conservation equation (\ref{covariant conservation}) corresponds to the conservation equations of the energy density (\ref{rhodotc}) and (\ref{rhodotX}). This turns the relation between covariant 4-vector $Q^\mu$ and the energy transfer $Q$ as
\begin{equation}
Q = Q_0.
\end{equation}

It is important to note that the coupling model (II) is not the expression of the covariant form due to the appearance of $H$. However, the covariant forms are introduced in order to provide the ability to compute the perturbations \cite{Maartens} and the coupling model (II) can provide the ability to compute the perturbations which has been performed in \cite{instable int dark}. Therefore, we allow the appearance of $H$ to arise in the coupling model (II).

Next we will consider the properties for each coupling form in detail by finding the fixed points and then investigating their stability. For simplicity, we will use the exponential potential ($V = V_{0}e^{-\eta x}$), the Gaussian potential ($V = V_{0}e^{-\eta x^{2}}$), and the inverse power law potential ($V = V_0 x^{-\eta}$) in our investigation. The evolution behavior is also investigated in order to compare with the standard evolution of the universe. Conveniently, we consider only the late time fixed points since they are relevant to the dynamics of dark energy. We also ignore the effect of baryon contribution in qualitative analysis because it has a tiny effect on the dynamical evolution comparable to other constituents. However, in the computational simulation of dynamical evolution, we add up all constituents in our consideration.

\subsection{Coupling model (I): $Q^\mu = \sqrt{2/3}\kappa\beta T^\tau_{\tau(c)} \left(1/24a^3\right)\epsilon^{\nu\rho\sigma\gamma}F_{\nu\rho\sigma\gamma} u^{\mu}$}

For this covariant form, the interaction term that satisfies equations (\ref{rhodotc}) and (\ref{rhodotX}) can be expressed as $Q = Q_{0} = \sqrt{2/3}\kappa\beta \rho_{c}(\dot{X} + 3HX)$ and the interaction variable becomes $\overline{\gamma} = \beta v^{2}$. The autonomous system in (\ref{auto2}) and (\ref{auto3}) can be rewritten as
\begin{eqnarray}
y' &=& \beta v^2 -\frac{3}{2}\lambda(x)z^{2}(x y -\sqrt{\frac{2}{3}})+\frac{3}{2}(w^{2}+\frac{4}{3}u^{2}+v^{2})y,\label{yint1}\\
v' &=& -\beta y v -\frac{3}{2}v[1+\lambda(x)xz^{2}-w^{2}-\frac{4}{3}u^{2}-v^{2}].\label{vint1}
\end{eqnarray}
The crucial properties of the fixed points for this interaction are summarized in Table \ref{table_int1}. Next we will discuss the results of each fixed point in detail.
\begin{table}[htdp]
\begin{center}\begin{tabular}{|c|c|c|c|c|c|c|}\hline
Point &  $y_*$ & $v_*$ & $\omega_{\text{tot}}$  & Acceleration?& Existence? & stability?\\\hline
& &  & &  &  &\\
A1, A2  & $-\frac{2}{3}\beta$ &$ \pm \sqrt{1-\frac{4}{9}\beta^2}$ & $-\frac{4}{9}\beta^2$  & $ |\beta| >\frac{\sqrt{3}}{2}$ & $|\beta|\leqslant \frac{3}{2}$ & $|\beta| > \frac{3}{2}$ \\
& &  & &  & & \\\hline
& &  & &  &  & \\
B1, B2 & $\pm1$ & $0$ & -1  & all $\beta$,$\lambda$ & all $\beta$,$\lambda$& $ \beta > - \frac{3}{2}$ \\

& &  & &  &  & \\\hline
& &  & &  &  & \\
C1, C2 & $-\frac{3}{2}\frac{1}{\beta}$ & $\pm \sqrt{-\frac{\sqrt{3/2}(4\beta^2-9)\lambda}{2\beta^2(2\beta-\sqrt{6}\lambda)}}$  & -1 & all $\beta$,$\lambda$  & Shown in & Shown in \\
& & & &  & Figure \ref{int1C}(a) & Figure \ref{int1C}(a) \\
& &  & &  &  &\\\hline
\end{tabular}
\caption{The properties of the fixed points for the coupling model (I) with the exponential potential. Note that $x = \sqrt{2/3}y$ at the fixed points.}\label{table_int1}
\end{center}
\end{table}

\begin{itemize}
\item
 {\bf Fixed points A1 and A2}
\end{itemize}
These fixed points correspond to $z = 0$. Since $\omega_{\text{tot}}$ depends on parameter $\beta$, it is easy to show that accelerating phase occurs when $|\beta| > \sqrt{3}/2$. Notice that the phantom phase $\omega_{\text{tot}} < -1$ is impossible because this will cause $v_{*}$ to be imaginary. The range of $\beta$ is constrained by the existence condition $|\beta| \leq 3/2$. The stability of these fixed points can be analyzed by considering the eigenvalues from the first order perturbation around each fixed point of the autonomous system (equation (\ref{dynamics1}), (\ref{yint1}), and (\ref{vint1})). These fixed points will be stable if the real part of all eigenvalues are negative. For these fixed points, the eigenvalues do not depend on the potential and read $\left(-3, \left(9 - 4\beta^{2}\right)/6, \left(9 - 4\beta^{2}\right)/3\right)$. This yields the stability condition $|\beta| > 3/2$. Since the stability condition is not compatible with the existence condition, we conclude that A1 and A2 are not stable. We are not interested in these fixed points.

\begin{itemize}
\item
 {\bf Fixed points B1 and B2}
\end{itemize}
\begin{figure}[htp]
\centering
\includegraphics[width=1.0\textwidth]{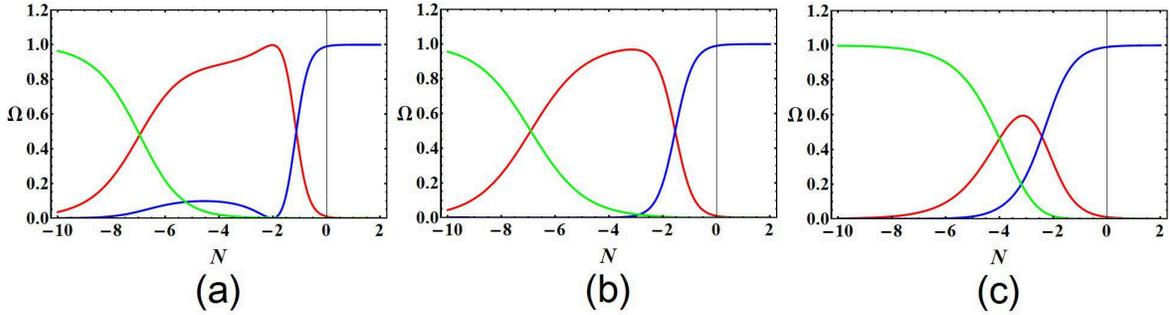}
\caption[$\rho_{cas}$]{This shows the evolution of the dynamical variables with different interaction parameter ($\beta$) at fixed point B, $y_{*} = +1$. The red line represents the energy parameter of dark matter ($\Omega_{c}$), the blue line represents the energy parameter of dark energy ($\Omega_{X}$) and the green line represents the energy parameter of radiation ($\Omega_r$). In this plot, we use the exponential potential with $\lambda = 1$ and $\beta = 0.5$ for (a), $\beta = 0$ for (b), and $\beta = -0.5$ for (c).}\label{evoint1_BP1}
\end{figure}
These fixed points correspond to dark energy dominated point and require $z = 0$. We cannot use these fixed points to solve coincidence problem because $\Omega_{X}/\Omega_{c}$ is not order of unity. However, it is instructive to investigate the effect of the coupling term on the dynamical evolution. The most important thing of this coupling form is that one can reproduce the non-interacting approach. There are no boundaries for the existence condition. For the stability condition, the eigenvalues do not depend on the potential and can be expressed as $(-3, 0, -3/2 - \beta)$. Thus, it guarantees that the fixed points are not stable for $\beta < -3/2$. In the case of $\beta > -3/2$, due to the zero eigenvalue,  we have to use the second order perturbations for analyzing the stability condition as seen in \cite{Koivisto} or use numerical method as we use here.

In numerical method, we check whether the evolution of the dynamical variables is convergent and provides the consistent behavior. For the exponential potential, the result is that if $y_{*} = 1$, the potential parameter needs to be positive ($\eta > 0$) and if $y_{*} = -1$, the potential parameter needs to be negative ($\eta < 0$). In other words, it is only allowed for $\eta y_{*} > 0$. From the first term in equation (\ref{vint1}), the transfer energy also depends on the sign of $y_{*}$. The result is that $\beta y_{*} > 0 $ corresponds to transfer of dark matter to dark energy and $\beta y_{*} < 0$ corresponds to transfer of dark energy to dark matter. However, if we carefully consider the evolution behavior of the dynamical variables, we will see that the interaction parameter ($\beta$) cannot take more value. This is due to the fact that the more value of the interaction parameter, the more change of the behavior of the variables at the early time. Figure \ref{evoint1_BP1} shows the effect of the interaction on energy density parameter evolution. We use the exponential potential with $\beta = 0.5$ in Figure \ref{evoint1_BP1}(a), $\beta = 0$ in Figure \ref{evoint1_BP1}(b), and $\beta=-0.5$ in Figure \ref{evoint1_BP1}(c). For these fixed points, they are allowed for both energy transfer from dark energy to dark matter and dark mater to dark energy.

The evolution for the Gaussian potential is similar to the exponential potential. The difference is that it is only stable for the $\eta > 0$ case. This result agrees with the analytic version in \cite{Koivisto}. For the inverse power law potential, the evolution is stable if $\eta \gtrsim -1$. It is consistent to the qualitative analysis as we have analyzed before.

\begin{figure}[htp]
\centering
\includegraphics[width=0.8\textwidth]{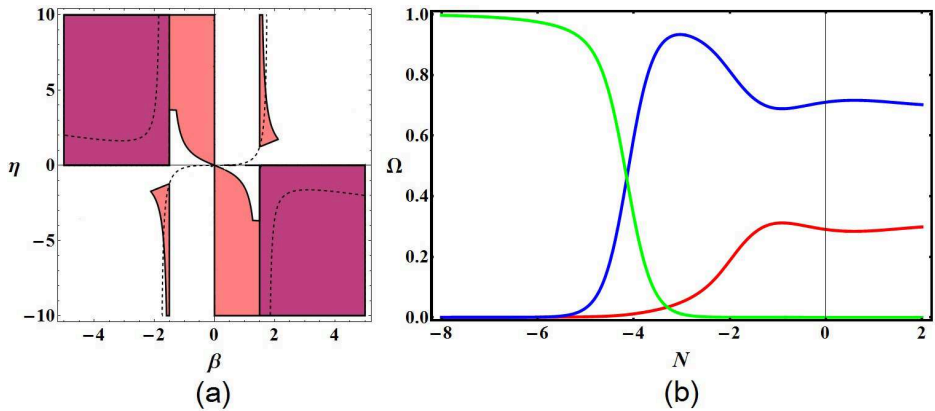}
\caption[$\rho_{cas}$]{(a): This figure shows the region of stability (red, shaded) and existence (blue, shaded) for the exponential potential in the $(\beta, \eta)$ parameter space. The violet region indicates the compatible region of two conditions. Dashed line represents a point at which the energy parameter ratio of dark energy to dark matter is 7:3. (b): This figure shows the evolution of the dynamical variables for the exponential potential. We use $\beta = -2.13$ for this simulation. The red line represents the energy parameter of dark matter ($\Omega_{c}$), the blue line represents the energy parameter of dark energy ($\Omega_{X}$) and the green line represents the energy parameter of radiation ($\Omega_{r}$).}\label{int1C}
\end{figure}

\begin{itemize}
\item
 {\bf Fixed points C1 and C2}
\end{itemize}

These fixed points correspond to dark energy dominated period at which  $\omega_{\text{tot}} = -1$. Moreover, these fixed points provide us the scaling regime. However, by considering the allowed region of the parameters for the stability and existence conditions, it cannot provide the well-behaved evolution as we will show now.

By using the same step, the plots of stability and existence region for the exponential potential are shown in the Figure \ref{int1C}(a). The dashed line, which shows the points at which the ratio of dark energy to dark matter is 7:3, is in the region that satisfies two conditions. In order to compare the result of this model with the observational data and solve the coincidence problem, $\beta$ and $\eta$ must satisfy the relation
\begin{equation}
\eta = \frac{2\sqrt{6}\beta^{3}}{45 - 14\beta^{2}}.
\end{equation}
This relation suggests that $\eta$ is divergent at $\beta = \pm\sqrt{45/14}$ as shown in Figure \ref{int1C}(a). This divergence also exists in the Gaussian potential and the inverse power law potential. From Figure \ref{int1C}(a), it is found that we will solve the coincidence problem by using these fixed points as long as $|\beta|>\sqrt{45/14}$. Due to the more value of the coupling parameter, it is obvious that the evolution at the early time will not satisfy the standard evolution of the universe as shown in Figure \ref{int1C}(b). We will see that there is no matter-dominated period in the evolution. The disappearance of matter-dominated period is also found for the Gaussian potential and the inverse power law potential.

Finally, it is important to note that there is the cosmological constant fixed point, $\lambda = 0$, like non-interacting case. The interaction terms do not affect this fixed point because the interaction terms depend on $v$ and $v = 0$ at this fixed point. Properties of this fixed point are similar to the non-interacting case. The difference is only that the matter does not completely dominate at matter-dominated period.

\subsection{Coupling model (II): $Q^\mu =\alpha H T^\tau_{\tau(c)} u^{\mu}$}

The interaction term that satisfies equations (\ref{rhodotc}) and (\ref{rhodotX}) of this coupling model can be expressed as $Q = Q_{0} = \alpha H\rho_{c}$ and the interaction variable becomes $\overline{\gamma} = \alpha v^{2}/2y$. The autonomous system in (\ref{auto2}) and (\ref{auto3}) can be rewritten as
\begin{eqnarray}
y' &=& \frac{1}{2} \frac{\alpha v^2}{y}-\frac{3}{2}\lambda(x)z^{2}(x y-\sqrt{\frac{2}{3}})+\frac{3}{2}(w^{2}+\frac{4}{3}u^{2}+v^{2})y,\label{auto2int2}\\
v' &=& -\frac{1}{2}\alpha v-\frac{3}{2}v[1+\lambda(x)xz^{2}-w^{2}-\frac{4}{3}u^{2}-v^{2}].
\end{eqnarray}
The properties of each fixed point are summarized in Table \ref{table_model2}. Note that there are the other fixed points at which the interaction parameter is $\alpha = -3$. It is not an interesting fixed point due to the much more value of $\alpha$. As we discussed in the previous interaction, the more value of $\alpha$, the more deviation from standard evolution of the universe in the past. Moreover, this value provides the non-dynamical evolution of the dark matter energy density, $\dot{\rho_{c}} = 0$. This behavior is not accepted for the observational data.

\begin{table}[htdp]
\begin{center}\begin{tabular}{|c|c|c|c|c|c|c|}\hline
Point &  $y_*$ & $v_*$ & $\omega_{\text{tot}}$  & Acceleration?& Existence? & stability?\\\hline
& &  & &  & & \\
A1, A2  & $\sqrt{-\frac{\alpha}{3}}$ & $\pm\sqrt{1+\frac{\alpha}{3}}$ & $\frac{\alpha}{3}$ & $\alpha < -1$ & $-3\leqslant\alpha\leqslant 0$ & $\alpha <-3$ \\
& &  & &  & &  \\\hline
& &  & &  &  & \\
B1, B2 & $\pm1$ & $0$ & -1 & all $\alpha$,$\lambda$ & all $\alpha$,$\lambda$  & $-3 < \alpha $ \\
& &  & &  &  &  \\\hline
\end{tabular}
\caption{The properties of the fixed points for the coupling model (II) with the exponential potential. Note that $x_* = \sqrt{2/3}y_*$ at the fixed points.}\label{table_model2}
\end{center}
\end{table}

\begin{itemize}
\item
 {\bf Fixed points A1 and A2}
\end{itemize}

These fixed points correspond to $z = 0$ and yield the phantom phase when $\alpha < -3$. However, this phase does not exist since $v$ is not real number. The existence condition for these fixed points is the same for all potentials. This condition can be expressed as $-3 \leq \alpha \leq 0$. To find the stability condition, we do the linear perturbations around the fixed points. The eigenvalues of these fixed points  read ($-3, \alpha + 3, \alpha + 3$). We will see that they do not depend on the potential form. Thus the stability condition is expressed as $\alpha < -3$. This condition conflicts with the existence condition for all potentials. This means that these fixed points are not stable for all potentials. We are not interested in these fixed points.

\begin{itemize}
\item
 {\bf Fixed points B1 and B2}
\end{itemize}

\begin{figure}[htp]
\centering
\includegraphics[width=1.0\textwidth]{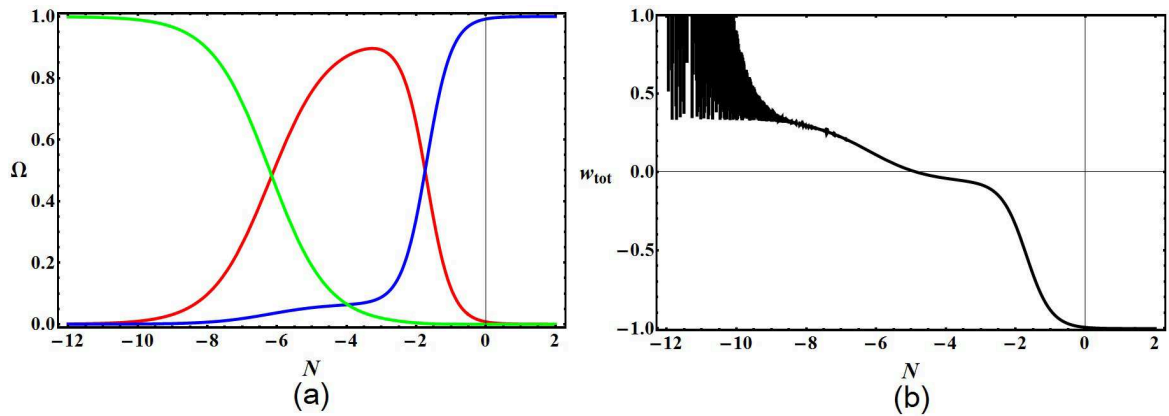}
\caption[$\rho_{cas}$]{(a) shows the evolution of the dynamical variables for $y_{*} = +1$ fixed point with the exponential potential. We use $\alpha = -0.2$, $\eta = 1.0$. The red line represents the energy parameter of dark matter ($\Omega_{c}$), the blue line represents the energy parameter of dark energy ($\Omega_{X}$) and the green line represents the energy parameter of radiation ($\Omega_{r}$). (b) shows the evolution behavior of the total equation of state parameter, $w_{\text{tot}}$, by using the same parameters with (a).}\label{evoint2_BP1}
\end{figure}

These fixed points correspond to dark energy dominated point without the effect of potential, $z=0$. These fixed points cannot be the points at which the coincidence problem is solved because of $\Omega_X = 1$. However, we will investigate the properties of these fixed points. These fixed points satisfy the existence condition for all points in parameter space ($\alpha, \eta$). This condition is valid for all potentials.

Considering the stability condition, the eigenvalues of the linear perturbation system do not depend on the potential form and read $(-3, 0, -3/2 - \alpha/2)$. As we discussed in the coupling model (I), a zero of the eigenvalues indicates that one cannot find the stability condition of these fixed points at the linear order perturbations. We check this by using a numerical method as we have done in the coupling model (I). Interestingly, it has similar properties to the coupling model (I). From the above eigenvalues, the stability condition is $\alpha > -3$. The crucial property of this coupling depends on the sign of $\alpha$. For the case of $\alpha < 0$ with the exponential potential,  $y_{*} = +1$ fixed point is stable when $\eta > 0$; conversely, $y_{*} = -1$ fixed point is stable when $\eta < 0$. In other words, they will be stable if $\eta y_{*} > 0$. For the Gaussian potential and the inverse power law potential, both $y_{*} = \pm 1$ fixed points will be stable if $\eta > 0$ and $\eta \gtrsim -1$ respectively.

\begin{figure}[htp]
\centering
\includegraphics[width=1.0\textwidth]{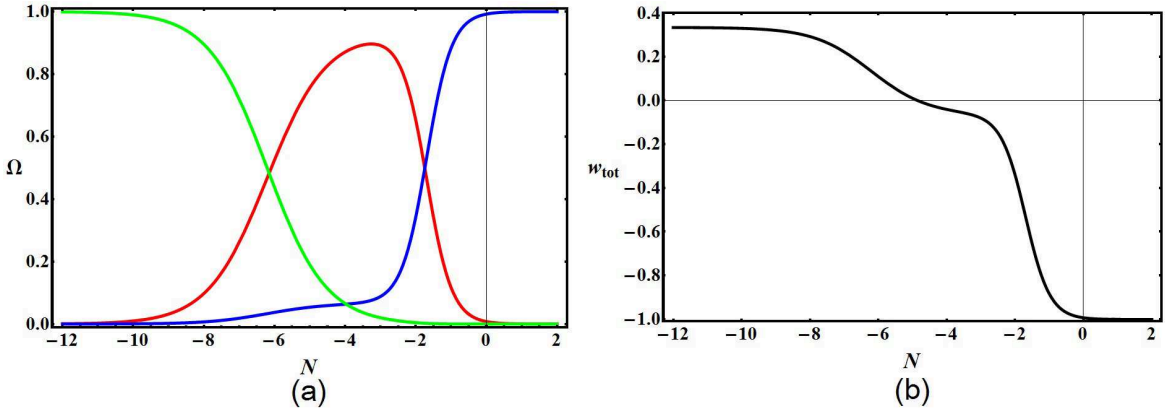}
\caption[$\rho_{cas}$]{(a) shows the evolution of the dynamical variables for $y_{*} = +1$ fixed point with the inverse power law potential. We use $\alpha = -0.2$, $\eta = 1.0$. The red line represents the energy parameter of dark matter ($\Omega_{c}$), the blue line represents the energy parameter of dark energy ($\Omega_{X}$) and the green line represents the energy parameter of radiation ($\Omega_{r}$). (b) shows the evolution behavior of the total equation of state parameter, $\omega_{\text{tot}}$, by using the same parameters with (a).}\label{evoint2_BP3}
\end{figure}

For the case of $\alpha > 0$, corresponding to energy transfer from dark matter to dark energy, with the exponential potential, we find that the evolution from numerical simulation encounters a singularity at the matter-dominated era. This singularity can be seen from the interaction term of dynamical equation (\ref{auto2int2}) when $v \neq 0$ and $y = 0$. However, this singularity does not exist in the case of $\alpha < 0$ as shown in Figure \ref{evoint2_BP1}. From this figure, we will see that $\Omega_{X}$ does not completely vanish at the matter-dominated period. This behavior also exists in this class of coupling form for the quintessence model as seen in \cite{pavon_new} and references therein. The physical interpreting of this behavior is that the energy density of dark matter is higher in value in the past than the present value of the energy density. This leads to the point that $v \rightarrow 1$ and $y \rightarrow 0$. Moreover, this coupling form also provides a clue to violate the second law of thermodynamics similar to the quintessence model \cite{violate second law}. Note that this singularity also exists in the Gaussian potential and the inverse power law potential when $\alpha > 0$.

From Figure \ref{evoint2_BP1}(b), we will see that $\omega_{\text{tot}}$ does not vanish at the matter-dominated period, $N \sim -3$, since $\Omega_c$ does not completely dominate. It is also found that $\omega_{\text{tot}}$ oscillates at the radiation-dominated period. This effect also occurs in all coupling models we investigate. To see how this effect can occur, we will return to the general form of the total equation of state parameter,
\begin{equation}
\omega_{\text{tot}} = -1 + (\gamma_r u^2 + \gamma_c v^2 + \gamma _b w^2 - \lambda x z^2).\label{wtot_gen}
\end{equation}
From this general form, we will see that $\omega_{\text{tot}} = \omega_{r}$ at the radiation-dominated period if $z$ completely vanishes. However, according to simulation computing, $z$ does not completely vanish. It oscillates in order of $10^{-9}$. As we have mentioned before, $x$ is a large value at early time. Thus, the large value of $x$ will factor the last term in equation (\ref{wtot_gen}) to oscillate in order of the huge value. However, for the inverse power law potential, $x$ will be eliminated by $\lambda = \eta/x $  and the oscillating behavior will disappear as shown in Figure \ref{evoint2_BP3}(b). From this figure, it is found that the oscillating behavior disappears and $\omega_{\text{tot}} \rightarrow \omega_r$ at radiation-dominated period as inferred from equation (\ref{wtot_gen}). This is the advantage of inverse power law potential. Thus, the inverse power law potential is favored by our numerical calculation and it implies that it is the most useful potential among all potentials we have investigated. This argument will hold only in the background analysis and may not be valid when the perturbation analysis is taken into account as suggested in \cite{Koivisto}. We leave this work in further investigation since it is beyond the purpose of our study.

The cosmological constant fixed point, $\lambda = 0$, also exists in this coupling model. The oscillating behavior also disappears since the last term in (\ref{wtot_gen}) completely vanishes. The well-behaved evolution of the dynamical variables is allowed only for $\alpha < 0$, corresponding to the energy transfer from dark energy to dark matter. The physical interpreting of this property is the same as we have mentioned above.

\subsection{Coupling model (III): $Q^\mu = \Gamma T^\tau_{\tau(c)} u^{\mu}$}

\begin{table}[htdp]
\begin{center}\begin{tabular}{|c|c|c|c|c|c|c|c|}\hline
Point &  $y_*$ & $v_*$ & $s_*$ & $\omega_{\text{tot}}$  & Acceleration?& Existence? & stability?\\\hline
& &  & &  & &  &\\
A1, A2  & $\sqrt{-s_* \gamma / 3}$ & $\pm\sqrt{1 - y_{*}^{2}}$ & any & $\frac{\gamma s_{*}}{3}$ & $\gamma s_{*} < 1$ & $-3 \leqslant \gamma s_{*} \leqslant 0$ & $\gamma s_{*} < -3$ \\
& &  & &  & & & \\\hline
& &  & &  & & &\\
B1, B2 & $\pm1$ & 0& any  & -1 & all $\gamma$,$\lambda$  & all $\gamma$,$\lambda$ & $\gamma s_{*} > -3$ \\
& &  & &  & & &\\\hline
& &  & &  & & &\\
C1, C2 & $0\leqslant y_*\leqslant 1$ & Eq(\ref{int3Crela1})& $-\frac{3}{\gamma}$ & -1 & all $\gamma$,$\lambda$  & all $\gamma$,$\lambda$ & - \\
& &  & &  & & &\\\hline
\end{tabular}
\caption{The properties of the fixed points for the coupling model (III) with the exponential potential. Note that $x_* = \sqrt{2/3}y_*$ at the fixed points.}\label{table_int3}
\end{center}
\end{table}
The interaction term that satisfies equations (\ref{rhodotc}) and (\ref{rhodotX}) of this coupling model can be expressed as $Q = Q_{0} = \Gamma \rho_{c}$. We cannot eliminate the Hubble parameter from the equation system of the dimensionless variables. We have to introduce new dimensionless variable for this coupling model. This variable has been introduced as $s = H_{0}/(H + H_{0})$ \cite{Maartens}. This form of this variable satisfies the compactness property but provides the singularity in the equation system at $s = 1$. In our consideration, we ignore the compactness property and choose the form of this variable for the simplicity of calculation. This new variable is expressed as
\begin{equation}
s \equiv \frac{H_{0}}{H},
\end{equation}
where $H_{0}$ is the Hubble parameter at the present time. Due to the fact that $H \propto 1/t$, it implies that $s \rightarrow 0$ at the early time and $s = 1$ at the present time. The interaction variable becomes $\overline{\gamma} = \gamma sv^{2}/2y$, where $\gamma = \Gamma/H_{0}$. The autonomous system in (\ref{auto2}) and (\ref{auto3}) can be rewritten as
\begin{eqnarray}
y'&=&\frac{\gamma sv^{2}}{2y}-\frac{3}{2}y[\lambda(x)xz^{2}-\gamma_{b}w^{2}-\gamma_{r}u^{2}-\gamma_{c}v^{2}]+\frac{\sqrt{6}}{2}\lambda(x)z^{2},\label{y primeint3}\\
v'&=&-\frac{\gamma s v}{2}-\frac{3}{2}v[\gamma_{c}+\lambda(x)xz^{2}-\gamma_{b}w^{2}-\gamma_{r}u^{2}-\gamma_{c}v^{2}],\label{v primeint3}
\end{eqnarray}
and the equation of new dimensionless variable can be expressed as
\begin{eqnarray}
s'&=&-\frac{3}{2}s[\lambda(x)x z^{2}-\gamma_{b}w^{2}-\gamma_{r}u^{2}-\gamma_{c}v^{2}].\label{s prime}
\end{eqnarray}
As we have seen, $s$ does not need to be stable at the present or at the early time due to the fact that it scales as $1/H$. It is well known that the Hubble parameter is not stable in all era of the late-time history of the universe (except inflationary era). However, we will consider in both cases, considering a fixed point of $s$ and ignoring it. The properties of each point  are summarized in Table \ref{table_int3}.

\begin{itemize}
\item
 {\bf Fixed points A1, A2}
\end{itemize}
These fixed points are $(x, y, v, w, u, s) = (0, \sqrt{-s_{*}\gamma/3}, \pm\sqrt{1 + s_{*}\gamma/3}, 0, 0, s_{*})$. This is the fixed points at which $s$ does not need to be a fixed point. If we require the fixed point of $s$, it turns out that $s = 0$. These fixed points can be reduced to $(x, y, v, w, u, s)=(0, 0, \pm1, 0, 0, 0)$ and corresponds to dark matter dominated solution. Its eigenvalues are ($-3, 3, 1/2$) for all the potentials. Therefore, these fixed points are not stable. However, if we return to ignore the fixed point of $s$, the eigenvalues will be ($-3, 3 + \gamma s, 3 + \gamma s$). Thus, the stability condition can be written as $\gamma s < -3$. This condition conflicts with existence condition, $-3 \leqslant \gamma s \leqslant 0$, and implies that these fixed points are not stable.
\begin{figure}[htp]
\centering
\includegraphics[width=1.0\textwidth]{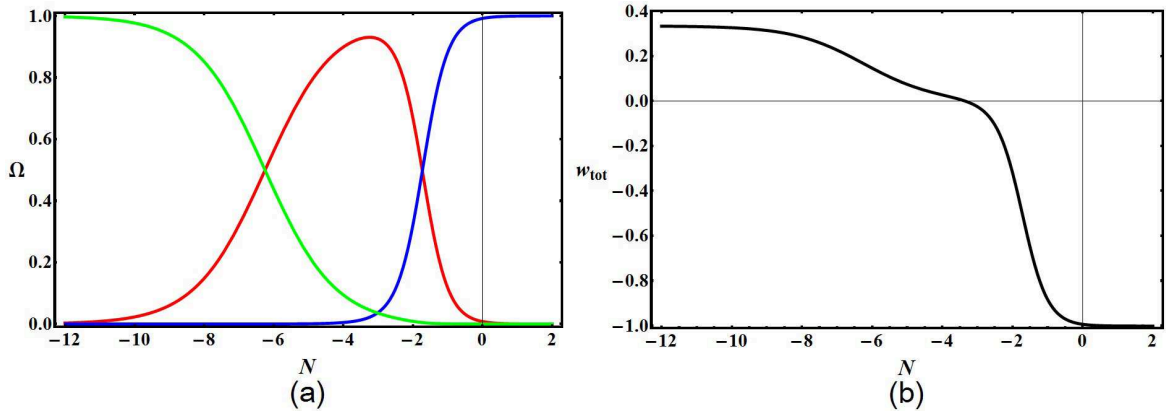}
\caption[$\rho_{cas}$]{(a) shows the evolution of the dynamical variables for $y_{*} = +1$ fixed point with the inverse power law potential. We use $\gamma = -0.2$, $\eta = 1.0$. The red line represents the energy parameter of dark matter ($\Omega_{c}$), the blue line represents the energy parameter of dark energy ($\Omega_{X}$) and the green line represents the energy parameter of radiation ($\Omega_{r}$). (b) shows the evolution behavior of the total equation of state parameter, $\omega_{\text{tot}}$, by using the same parameters with (a).}\label{evoint3_BP3}
\end{figure}
\begin{itemize}
\item
 {\bf Fixed point B1, B2}
\end{itemize}
These fixed points are $(x, y, v, w, u, s) = (\pm\sqrt{2/3}, \pm1, 0, 0, 0, s_{*})$. They correspond to three-form dominated solution and their properties are the same for all potentials. It gives the total equation of state parameter like the cosmological constant, $\omega_{\text{tot}} = -1$. The eigenvalues of these fixed points are ($-3, 0, (-3 - \gamma s_{*})/2$). Therefore, we can infer nothing about its stability from the linear analysis because of the eigenvalue 0. However, the last eigenvalue suggests us that the stability condition requires at least $\gamma s_{*} > -3$. If this condition is satisfied, the stability condition will be investigated by using numerical method as done in the previous coupling models. It turns out that these fixed points are stable. The crucial properties of these fixed points are similar to the coupling model (II). The well behavior of the energy density evolution is allowed only for negative sign of interaction parameter, $\gamma < 0$. In other words, it is allowed for energy transfer from dark energy to dark matter only. The physical interpretation of this phenomenon is the same as mentioned in the coupling model (II).

The advantage of this coupling form is that it has a tiny effect on the evolution behavior at the matter- and radiation-dominated periods as shown in Figure \ref{evoint3_BP3}. From this figure, the evolution behavior at the matter- and radiation-dominated periods is close to the standard evolution of the universe. This feature is explained by considering the interaction terms in equations (\ref{y primeint3}) and (\ref{v primeint3}). These interaction terms effectively vanish since $s$ approaches zero during matter- and radiation-dominated periods. They also have a tiny effect at the future time since $v \rightarrow 0$ for these fixed points. This keeps the stability of these fixed points for all variables except $s$ even though $s$ still evolves.

\begin{itemize}
\item
 {\bf Fixed point C1, C2}
\end{itemize}
These fixed points are $(x, y, v, w, u, s) = (x_{*}, y_{*}, v_{*}, 0, 0, s_{*})$, where $x_{*} = \sqrt{2/3}y_{*}$, $0\leqslant y_*\leqslant 1$,
\begin{equation}
v_{*} = \pm\sqrt{\frac{2\lambda y_{*}(1 - y_{*}^{2})}{\sqrt{6} + 2\lambda y_{*}}},\label{int3Crela1}
\end{equation}
and
\begin{equation}
\gamma s_{*} = -3.\label{int3Crela2}
\end{equation}
The most important properties of these fixed points are related to the second relation (\ref{int3Crela2}). This relation yields $\Gamma = -3H$. It provides not only the more value of the interaction parameter but also the non-dynamics of energy density of dark matter due to equation (\ref{rhodotc}). This gives the similar results to the case of $\alpha = -3$ in the coupling model (II) by replacing $\gamma s_{*}$  with $\alpha$. The difference is that $\gamma s_{*}$ varies with time and approaches zero at the early time while $\alpha$ is constant. However, the evolution behavior is not significantly different since the matter will run out before it dominates in the past due to the more value of $\gamma s_{*}$ at the fixed points.

For the case of $s$ to be a fixed point, it is found that $s_* = 0$ and this yields the infinity of the coupling parameter, $\gamma =-3/s_{*} \rightarrow \infty$. If we take a small perturbation by setting $s_* = \varepsilon$ where $\varepsilon$ is a small value, it turns out that this fixed points are not stable and evolve to fixed points B.

Finally, we note that there is the cosmological constant fixed point, $\lambda = 0$, similar to coupling model (I) and (II). The difference is that the evolution behavior are tinily affected from the interaction since $s \rightarrow 0$ at the matter-dominated period. We also found that one can set $s$ as any value at this stable fixed point due to $v = 0$ fixed point.

\section{Extension to Another Coupling Model}\label{extension}

We will begin this section by summarizing the results of three coupling models.
\begin{itemize}
\item
 {Coupling model (I)}
\end{itemize}
There are three classes of all fixed points. The A1 and A2 fixed points are not stable. For the B1 and B2 fixed points, they are stable but one cannot use them to solve coincidence problem since they are the fixed points at which $y= \pm 1$. However, it is instructive to consider this fixed points in order to study the effect of the interaction because it can be reduced to the non-interacting approach. The energy can transfer from dark energy to dark matter and dark matter to dark energy. Finally, for fixed points C1 and C2, the coupling parameter is allowed only in region with large value. Therefore, there is no matter-dominated period for these fixed points.

\begin{itemize}
\item
 {Coupling model (II)}
\end{itemize}
Actually, there are three classes for this coupling model, but we show only two classes because the missing one provides the much more value of the coupling parameter, $\alpha = -3$. For the fixed points A1 and A2, they are not stable like the fixed points A1 and A2 in the coupling (I). The coincidence problem cannot be solved by using the fixed points B1 and B2 because of $y =\pm 1$. The energy transfer is allowed only from dark energy to dark matter. According to the evolution behavior, we found that the inverse power law potential is suitable for the interacting three-form field.

\begin{itemize}
\item
 {Coupling model (III)}
\end{itemize}
The properties are very similar to the coupling model (II) although there are more variables in phase space. They are similar when we take $\gamma s_* \rightarrow \alpha$. This means that all properties are the same as the coupling model (II) except that the effective coupling $\gamma s_*$ now varies with time. $\gamma s_* $ is equal to $-3$ at the fixed points but not necessary hold for other points in phase space. This provides the vanishing value of effective coupling at the matter-dominated period because of $s \rightarrow 0$. However, the evolution behavior of these fixed points is not compatible with observation since there is no matter-dominated period. In other words, the matter will run out before it dominates in the past due to the more value of $\gamma s_*$ at the fixed points.

The cosmological constant fixed point, $\lambda = 0$, also exists in all coupling models we have investigated. It is the stable fixed point with $\rho_X = \text{constant}$ and $\omega_X = -1$. Therefore, the coincidence problem will not be solved by this type of the fixed point.

From summary above, there are no coupling models which provide the possibility to solve the coincidence problem. We now introduce the another coupling model in order to solve it. This coupling model can be written in the covariant form as
\begin{equation}
Q^\mu = -\frac{\sqrt{6}\Gamma}{\kappa} \frac{1}{24 a^3}\epsilon^{\nu\rho\sigma\gamma}F_{\nu\rho\sigma\gamma} u^{\mu},\label{int4}
\end{equation}
where $\gamma$ is the coupling parameter for this model. For this covariant form, the interaction term that satisfies equations (\ref{rhodotc}) and (\ref{rhodotX}) can be expressed as $Q = Q_{0} = \sqrt{6}\Gamma (\dot{X} + 3HX)/\kappa$. One cannot eliminate $H$ from the dimensionless equations as found in coupling model (III). We introduce new dimensionless variable $s = H_0/H$ as done in coupling model (III). We note that it is possible to eliminate $H$ from the dimensionless equations by introducing the covariant interaction form as
\begin{equation}
Q^\mu = -\frac{\sqrt{6}\gamma H^2}{\kappa} \frac{1}{24 a^3}\epsilon^{\nu\rho\sigma\gamma}F_{\nu\rho\sigma\gamma} u^{\mu}.\label{int5}
\end{equation}
However, it encounters the singularity at the radiation-dominated period. We will discuss this issue later.
The interaction term in equation (\ref{int4}) yields $\overline{\gamma} = \gamma s^{2}$, where $\Gamma = H^2_0 \gamma$. The dynamical equations associated with the interaction term become
\begin{eqnarray}
y' &=& \gamma s^{2}-\frac{3}{2}\lambda(x)z^{2}(x y-\sqrt{\frac{2}{3}})+\frac{3}{2}(w^{2}+\frac{4}{3}u^{2}+v^{2})y,\label{auto2int4}\\
v' &=& -\frac{\gamma s^{2} y}{v}-\frac{3}{2}v[1+\lambda(x)xz^{2}-w^{2}-\frac{4}{3}u^{2}-v^{2}],\label{auto3int4}\\
s' &=& -\frac{3}{2}s[\lambda(x)x z^{2}-\gamma_{b}w^{2}-\gamma_{r}u^{2}-\gamma_{c}v^{2}].\label{s primein4}
\end{eqnarray}
Note that the dynamical equations for the interaction (\ref{int5}) are obtained by ignoring the equation (\ref{s primein4}) and setting the interaction parameter as $\gamma s^{2} \rightarrow \gamma $. From the first term in equation (\ref{auto3int4}), we will see that the energy density of dark matter cannot vanish. This guarantees that there are no stable fixed points at which $\Omega_{c} = 0$ at the present time. It is also found that this term will provide the singularity at radiation-dominated period since $v \rightarrow 0$ for the interaction (\ref{int5}). However, for the interaction (\ref{int4}), the singularity can be avoided since $s \rightarrow 0$ at the radiation-dominated period. This mechanism is similar to a mechanism in the coupling model (III) in which the effect of the interaction is very small at early time. Therefore, the interaction in (\ref{int5}) is not interesting for us and we will not consider it. The properties of all fixed points are summarized in Table \ref{table_model4}. Conveniently, we introduce the new parameters as
\begin{eqnarray}
A &\equiv& \left(9\gamma s^{2}_* +3\sqrt{-3 + 9 (\gamma s^{2}_*)^2} \right)^{1/3},\\
B &\equiv& -(\gamma s^{2}_*)^2 -3\sqrt{6} \frac{ \gamma s^{2}_*}{\lambda}.
\end{eqnarray}

\begin{table}[htdp]
\begin{center}\begin{tabular}{|c|c|c|c|c|c|c|}\hline
Point &  $y_*$ & $v_*$  & $\omega_{\text{tot}}$ & Acceleration?& Existence? & Stability? \\\hline
& &  & &  &  &\\
A1, A2 & $\frac{1}{A} + \frac{A}{3}$ & $\pm\sqrt{1-y_*^2}$ & $-y_*^2$ & $-\frac{1}{\sqrt{3}}\leqslant\gamma$ & $-\frac{1}{\sqrt{3}}\leqslant\gamma\leqslant 0$  & $\gamma > 0$\\
& &  & &  & & \\\hline
& &  & &  &  &\\

A3, A4  & $-\frac{1+i\sqrt{3}}{2A}-\frac{(1-i\sqrt{3})A}{6}$ & $\pm\sqrt{1-y_*^2}$  & $-y_*^2$ & $\gamma\leqslant \frac{1}{\sqrt{3}}$ & $0\leqslant\gamma\leqslant \frac{1}{\sqrt{3}}$ & $\gamma < 0$ \\
& &  & &  &  & \\\hline
& &  & &  & & \\

A5, A6  & $-\frac{1-i\sqrt{3}}{2A}-\frac{(1+i\sqrt{3})A}{6}$ & $\pm\sqrt{1-y_*^2}$ & $-y_*^2$  & No & $|\gamma| \leqslant \frac{1}{\sqrt{3}}$ & $|\gamma| > \frac{1}{\sqrt{3}}$\\
& &  & &  &  & \\\hline
& &  & &  & & \\

B1, B2 & $\frac{\gamma}{3} + \frac{1}{3}\sqrt{9 - B}$ & $\pm\sqrt{-\frac{2}{3}\gamma y_*}$  & -1 & all $\gamma$,$\lambda$ & Figure \ref{int4Bp1}& Figure \ref{int4Bp1} \\
& &  & &  &  & \\\hline
& &  & &  & &  \\

B3, B4 & $\frac{\gamma}{3} - \frac{1}{3}\sqrt{9 - B}$ & $\pm\sqrt{-\frac{2}{3}\gamma y_*}$  & -1 & all $\gamma$,$\lambda$  & No & Figure \ref{int4Bp1} \\
& &  & &  &  &\\\hline
\end{tabular}
\caption{The properties of the fixed points for the coupling model (IV). Note that $x_{*} = \sqrt{2/3}y_{*}$ at the fixed points. We set $s_* = 1$ for convenience. To restore $s_*$ into the table, one just replaces $\gamma$ by $\gamma s^2_*$. For fixed point B, solutions of $y_*$ depend on the potential form. In this table, we show only for the solution of exponential potential. For the other potentials, we will discuss in the description of each potential. }\label{table_model4}
\end{center}
\end{table}

\begin{itemize}
\item
 {\bf Fixed points A1-A6}
\end{itemize}
These fixed points correspond to $z_* = 0$. The constraint equation for these fixed points can be written as
\begin{eqnarray}
y^3_*- y_*-\frac{2}{3}\gamma s^2_*= 0.
\end{eqnarray}
Three solutions of $y_*$ for this equation are expressed in Table \ref{table_model4}. The conditions for stability and existence of all A1-A6 do not depend on potential form. These two conditions are not compatible each other. We are not interested in all these fixed points.

\begin{itemize}
\item
 {\bf Fixed points B1-B4}
\end{itemize}
These fixed points correspond to the solutions with $z_* \neq 0$. One of the conditions for these fixed points is
\begin{eqnarray}
\lambda x_* z_*^{2}-\gamma_{b}w_*^{2}-\gamma_{r}u_*^{2}-\gamma_{c}v_*^{2}= 0.\label{constraintint4b2}
\end{eqnarray}
This condition leads to $\omega_{\text{tot}} = \omega_{X} = -1$ and also provides the fixed point $s_* \neq 0$. It is possible to set $s_* = 1$ as the fixed point at the present time to get $H = H_0$. The condition (\ref{constraintint4b2}) can be rewritten as
\begin{eqnarray}
y^2_*-\left(\frac{2}{3}\right)\gamma s^2_* y_*-\left(1+\sqrt{\frac{2}{3}}\frac{\gamma s^2_* }{\lambda }\right)= 0 \;\ \text{and} \;\
v^2_* = -\frac{2}{3}\gamma s^2_* y_*.\label{constraintint4b}
\end{eqnarray}

The solution of $y_*$ is strongly dependent on the potential form. There are two solutions for the exponential potential and inverse power law potential and three solutions for the Gaussian potential. From this condition, we will see that it is possible to get the fixed points at which $v^2_*/y^2_* \sim O(1)$ with $s_* = 1$. This allows us to solve the coincidence problem since $\Omega_{c0}/ \Omega_{X0}  = v^2_*/(y^2_* + z^2_*)$. Our investigation of these fixed points will be separated into three parts depending on the potential.

\begin{figure}[htp]
\centering
\includegraphics[width=0.8\textwidth]{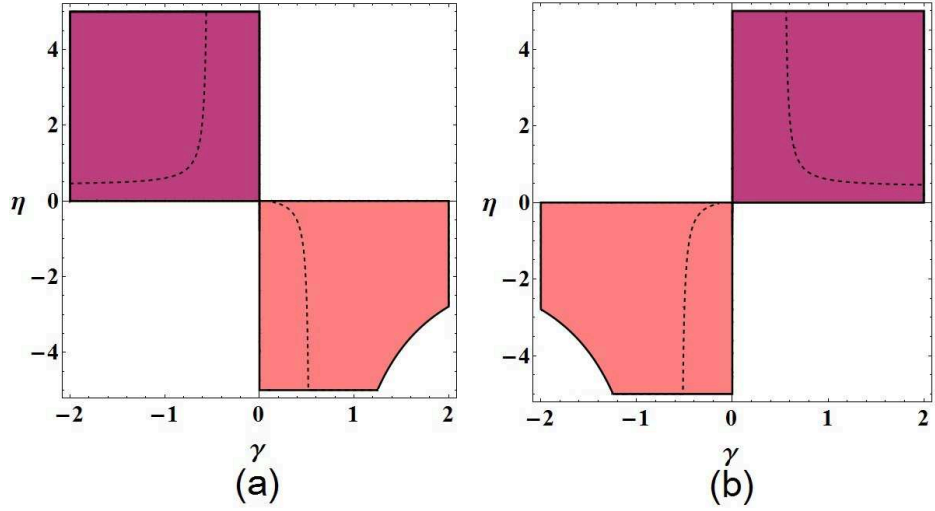}
\caption[$\rho_{cas}$]{This figure shows the region of stability (red, shaded) and existence (blue, shaded) in the $(\gamma,\eta)$ parameter space for the inverse power law potential $V = V_{0}x^{-\eta}$. The violet region indicates the compatible region of two conditions. Dashed line represents a point at which the energy parameter ratio of dark energy to dark matter is 7:3. Figure (a) and (b) correspond to $y_{*-}$ and $y_{*+}$ solution in equation (\ref{yBint4p3}). Note that we set $s_* = 1$ in this plot. However, we can restore $s_*$ into this plot by setting $\gamma \rightarrow \gamma s^2_*$.}\label{int4Bp3}
\end{figure}
\begin{itemize}
\item
  Inverse power law potential
\end{itemize}

For inverse power law potential, $\lambda = \eta/x_* = \sqrt{3/2} \eta/y_*$. By substituting $\lambda$ into equation (\ref{constraintint4b}), the solutions of this equation are
\begin{eqnarray}
y_{*\pm} &=& \frac{2 \gamma s^2_* +2 \gamma s^2_* \eta \pm \sqrt{36 \eta ^2+(-2 \gamma s^2_* -2 \gamma s^2_* \eta )^2}}{6 \eta }.\label{yBint4p3}
\end{eqnarray}

The conditions of existence and stability are shown in Figure \ref{int4Bp3}. The minus solution corresponds to the Figure \ref{int4Bp3}(a), and plus solution corresponds to the Figure \ref{int4Bp3}(b). By using the solutions in equation (\ref{yBint4p3}), one can find the relation of $\gamma, \eta, s_*$ and $\Omega_{c}$ as
\begin{eqnarray}
\eta = \frac{4\Omega_c (\gamma s^2_*)^2}{9\Omega^2_c - 4(1-\Omega_c)(\gamma s^2_*)^2}.\label{relaint4Bp3}
\end{eqnarray}

The characteristic properties of this relation are found in the Figure \ref{int4Bp3}. From this relation, the singularity of the curve $\Omega_{X}/\Omega_{c}$ in Figure \ref{int4Bp3} can be written as
\begin{eqnarray}
\gamma s^2_* = \pm \frac{3\Omega_c}{2\sqrt{1-\Omega_c}}.
\end{eqnarray}

For $\Omega_c = 0.3$, we get the singularity at $\gamma s^2_* \approx \pm 0.538$ as shown in Figure \ref{int4Bp3}. From Figure \ref{int4Bp3} and relation (\ref{relaint4Bp3}), It is also found that the stability condition can be written as
\begin{eqnarray}
|\gamma s^2_*| > \frac{3\Omega_c}{2\sqrt{1-\Omega_c}}.
\end{eqnarray}
This condition also suggests us that the strength of the interaction depends on the value of $\Omega_c$. If $\Omega_c$ increases, the lower bound of this condition also increases. Thus $\Omega_c$ cannot take large value since interaction will be too large.

\begin{figure}[htp]
\centering
\includegraphics[width=0.9\textwidth]{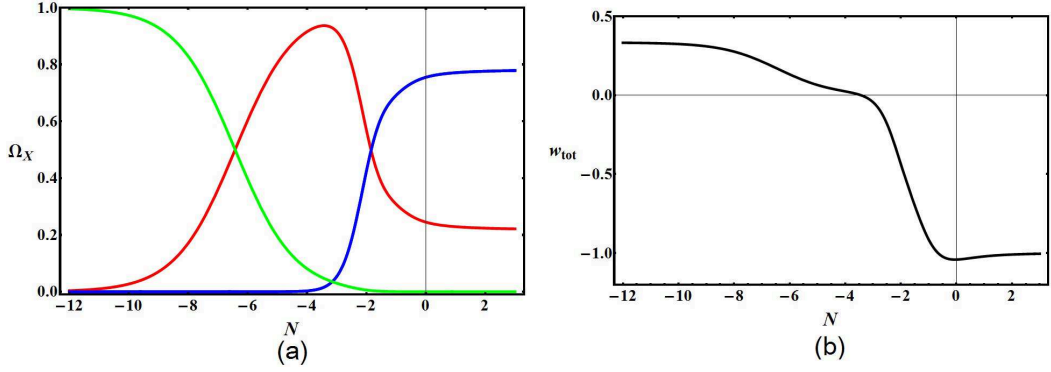}
\caption[$\rho_{cas}$]{(a) shows the evolution of the dynamical variables in extension the coupling model with the inverse power law potential. We use $\gamma = -0.5$, $\eta = \frac{16 \gamma ^2}{3 \left(-3+16 \gamma ^2\right)}$ and $s_* = 1$ for this simulation. The red line represents the energy parameter of dark matter ($\Omega_{c}$), the blue line represents the energy parameter of dark energy ($\Omega_{X}$) and the green line represents the energy parameter of radiation ($\Omega_{r}$). (b) shows the evolution behavior of the total equation of state parameter by using the same parameters with (a).}\label{evoint4Bp3}
\end{figure}

For the evolution behavior, we set $\gamma = -0.5$, $s_* = 1$ and choose $\Omega_c \sim 0.25$. This leads to $\eta = 16 \gamma ^2/3 \left(-3+16 \gamma ^2\right) \approx 1.333$. By using this parameter set, the evolution of energy density parameters and $\omega_{\text{tot}}$ are shown in Figure \ref{evoint4Bp3}. It is found that one can solve the coincidence problem by using this coupling model because there are the stable fixed points at which $\Omega_{X}/\Omega_{c} \sim 7/3$ and there are the matter- and radiation-dominated periods with the same parameters. Note that, in the simulation, we have to set a small deviation from the fixed points in order to get the dynamics of the variables. Therefore, the stable fixed points will be slightly deviated from the fixed point we set as found in Figure \ref{evoint4Bp3}.

\begin{figure}[htp]
\centering
\includegraphics[width=1.0\textwidth]{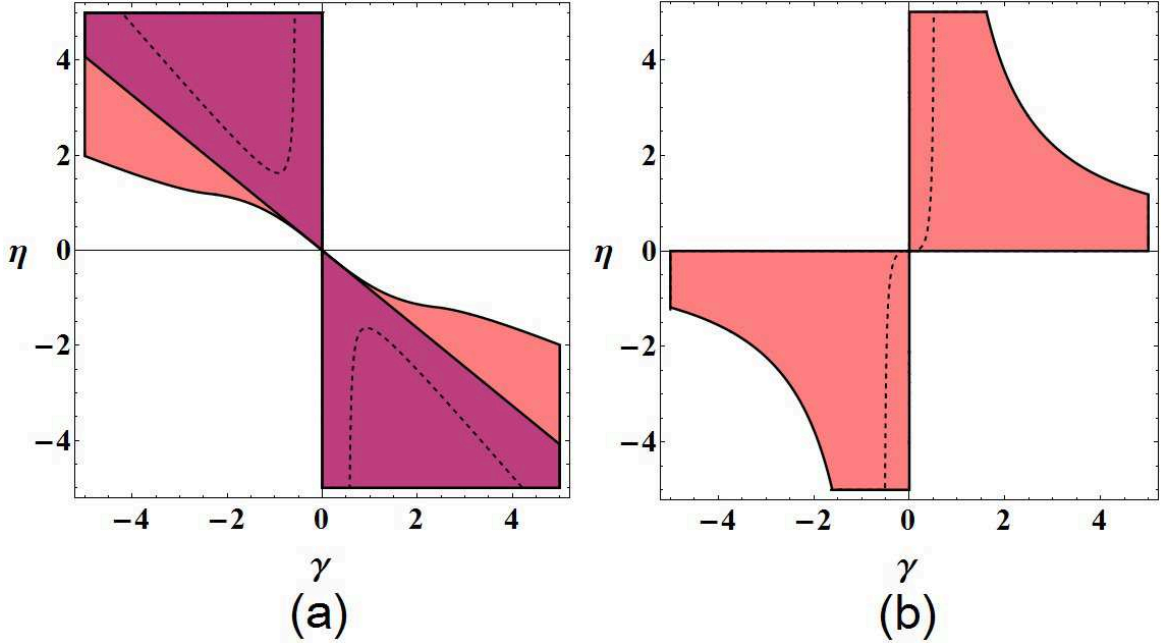}
\caption[$\rho_{cas}$]{This figure shows the region of stability (red, shaded) and existence (blue, shaded) in the $(\gamma,\eta)$ parameter space for the exponential potential $V = V_{0}e^{-\eta x}$. The violet region indicates the compatible region of two conditions. Dashed line represents a point at which the energy parameter ratio of dark energy to dark matter is 7:3. Figure (a) and (b) correspond to $y_{+}$ and $y_{-}$ solution respectively. Note that we set $s_* = 1$ in this plot. However, we can restore $s_*$ into this plot by setting $\gamma \rightarrow \gamma s^2_*$. }\label{int4Bp1}
\end{figure}

\begin{itemize}
\item
  Exponential potential
\end{itemize}
For the exponential potential, $\lambda = \eta$ and the solutions of $y_*$ corresponding to the equation (\ref{constraintint4b}) are
\begin{eqnarray}
y_{*\pm} &=& \frac{\gamma s^2_*\eta \pm \sqrt{\eta\left(3\sqrt{6}\gamma s^2_* + 9\eta + (\gamma s^2_*)^{2}\eta\right)}}{3\eta}.\label{yBint4p1}
\end{eqnarray}
The plus solution corresponds to the fixed points B1 and B2, and minus solution corresponds to the fixed points B3 and B4.  We show the existence and stability conditions by using numerical method as seen in Figure \ref{int4Bp1}. From this figure, only $y_{*+}$ solution is consistent while there are no existence regions for $y_{*-}$ solution. Thus, the interesting fixed points are B1 and B2 which have the same properties. For $\Omega_{X}/\Omega_{c} = 7/3$, the stable condition allows only the region in which $|\gamma| \geqslant \sqrt{81/280}\approx 0.538$.

The calculation and analysis are performed in the same step as done in the inverse power law potential. The results are similar to the inverse power law case. The difference is that the evolution behavior for exponential potential will encounter the large oscillating of $\omega_{\text{tot}}$ at radiation-dominated period. We will not show this evolution behavior here.

\begin{figure}[htp]
\centering
\includegraphics[width=1.0\textwidth]{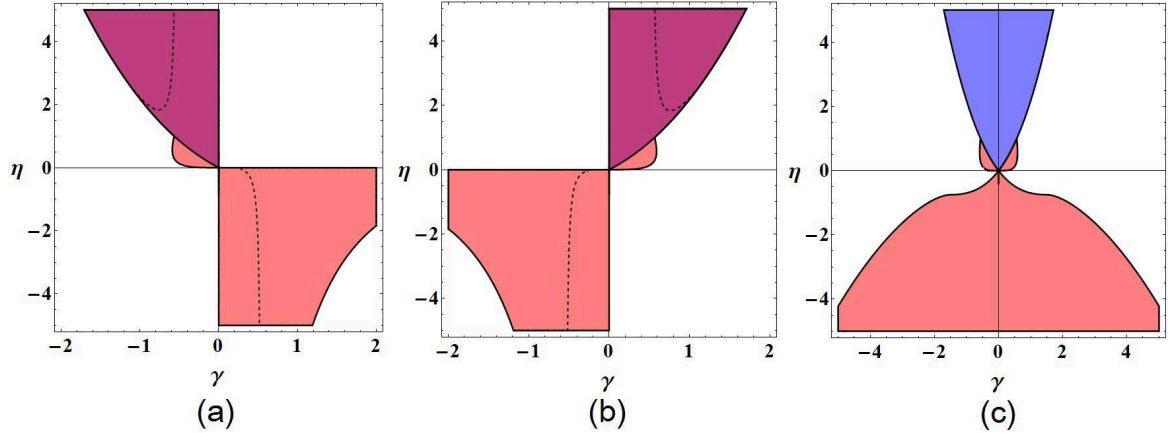}
\caption[$\rho_{cas}$]{This figure shows the region of stability (red, shaded) and existence (blue, shaded) in the $(\gamma,\eta)$ parameter space for the exponential potential $V = V_{0}e^{-\eta x}$. The violet region indicates the compatible region of two conditions. Dashed line represents a point at which the energy parameter ratio of dark energy to dark matter is 7:3. Figure (a), (b), and (c) correspond to three solutions of equation (\ref{constraintint4b}).}\label{int4Bp2}
\end{figure}
\begin{itemize}
\item
  Gaussian potential
\end{itemize}

For the Gaussian potential, $\lambda = 2\eta x_* = \sqrt{8/3}\eta y_*$. Equation (\ref{constraintint4b}) become
\begin{eqnarray}
y^2_*-\frac{2 y_* \gamma s^2_* }{3}-1-\frac{\gamma s^2_*}{2 \eta y_*  }= 0.\label{constraintint4bp2}
\end{eqnarray}

There are three solutions of this equation. We will not show the exact solutions here for convenience because they are lengthy. The region plot of these three solutions are shown in Figure \ref{int4Bp2}. The evolution behavior is similar to the inverse power law and exponential potential case. Thus, we will not show this simulation here.

Finally, it is important to note that there is no cosmological constant fixed point, $\lambda = 0$, similar to three coupling models we have investigated. This can be seen from the interaction terms of this coupling model, (\ref{auto2int4}) and (\ref{auto3int4}). These terms do not vanish when $v \rightarrow 0$. Therefore, it yields the results equivalent to the fixed points A which $z_* = 0$ and $v^2_* = -2 \gamma s^2_*/3 y_*$. It is not a stable fixed point for this coupling model.

\section{Conclusions and Discussions}\label{Conclusions}

The possibility of the three-form field to be dark energy is investigated. We begin this work by summarizing the results of three-form field dark energy. By analyzing the dynamical equations of the system, we have found that the potential for three-form field should be runaway. We have also found that three-form field with proper potential can act as dark energy. However, non-interacting three-form field cannot provide the possibility to solve the coincidence problem since there are no stable fixed points at which $\Omega_X/\Omega_c \sim O(1)$. Analogous to scalar field dark energy, we examine the interacting three-form field with dark matter and find the possibility to solve the coincidence problem. We introduce three covariant coupling forms which imitate idea from scalar coupling forms. Although the coupling model (II) can not be expressed in the covariant form due to the appearance of $H$, we can compute the perturbations for this coupling model as seen in \cite{instable int dark}. Therefore, we allow the appearance of $H$ to arise in this coupling model. We have found that three coupling forms cannot lead to the well-behaved solution of the coincidence problem. Therefore, we introduce the forth coupling form in order to provide the ability to solve the coincidence problem. The properties for each coupling model are summarized as follows.

For coupling (I), $Q = \sqrt{2/3}\kappa\beta \rho_{c}(\dot{X} + 3HX)$, the fixed points A are not stable. Their eigenvalues can be reduced to case of non-interaction. The effects of the interaction in the evolution of the energy density parameters are discussed in both qualitative and numerical analysis. We find the stability of the fixed points B numerically. By using three runaway potentials including exponential, Gaussian and inverse power law potentials, the results are that the fixed points B are stable and the stability is independent of sign of the coupling. However, they cannot solve the coincidence problem since they give the stable fixed points at which $\Omega_X = 1$. However, it is instructive to study the effect of interaction in this coupling model. This effect is examined and the result is shown in Figure \ref{evoint1_BP1}. For the exponential potential, $V = V_{0}e^{-\eta x}$, the standard evolution of the universe, which has the matter- and radiation-dominated period, is obtained when $y_{*} \eta > 0$. It corresponds to the runaway potential since the sign of $y$ is the same as the sign of $x$. For the inverse power law and Gaussian potentials, both $y_{*} = 1$ and $y_{*} = -1$ give the well-behaved evolution when $\eta \gtrsim -1$ and $\eta > 0$ respectively. For the fixed points C, they can make an explanation of the coincidence problem. However, the coupling parameter $\beta$ is of order 2, which makes the evolution at the early time different from the actual evolution as shown in the Figure \ref{int1C}(b).

For coupling (II), $Q = \alpha H \rho_{c}$, the fixed points A are not stable. Their eigenvalues cannot be reduced to case of non-interacting one. This is because there is $\alpha v^{2}/2y^{2}$ term which does not vanish when $\alpha = 0$. For the fixed points B, the stability depend upon sign of the coupling. The singularity appears in the case of $\alpha > 0$ at the matter-dominated period, but is not in the case of $\alpha < 0$ for all potentials. This is because $\alpha > 0$ corresponds to energy transfer from dark matter to dark energy. This means that the energy density of dark matter increases when we go back in time. During matter-dominated period, the value of energy density of dark matter must take the enough value in order to give the observed value at the present time. For this reason, the singularity is emerged when $v\rightarrow 1$ and $y\rightarrow 0$. On the other hand, $\alpha < 0$ corresponds to energy transfer from dark energy to dark matter. There is no a singularity in this case because the value of the energy density of dark matter does not need to be large in the past. This implies that the energy transfer is allowed from the dark energy to dark matter only. Moreover, there are the other fixed point corresponding to $\alpha = -3$. However, $\alpha$ is too large in value for the evolution to agree with the cosmological observations. Moreover, it gives non-dynamical evolution of the dark matter energy density.

We also show that the suitable potential for the well-behaved evolution of three-form dark energy is power law potential. This property can be seen by comparing the evolution behavior of $\omega_{\text{tot}}$ for exponential potential (Figure \ref{evoint2_BP1}) and inverse power law potential (Figure \ref{evoint2_BP3}). There is the oscillating behavior of $\omega_{\text{tot}}$ for exponential potential. We interpret this behavior from the last term in equation (\ref{wtot_gen}). The enormous value of $x$ will factor this term to dominate even though $z^2$ approaches to zero. It is important to note that the inverse power law potential may give an unstable evolution of the cosmological perturbations since it provides negative sound speed square \cite{Koivisto}. It is of interest to investigate whether the interaction can alleviate this behavior and we leave this investigation in further works.

For coupling (III), $Q = \Gamma \rho_{c}$, One cannot eliminate the Hubble parameter $H$ from the dynamical equations. We introduce new dimensionless variable in order to eliminate $H$ by $s \equiv H_0/H$. The results from this coupling are very similar to ones of the coupling (II), just $\alpha$ becoming $\gamma s$. The difference is that $\alpha$ is constant throughout the evolution of the universe, while $\gamma s$ varies with time. However, by comparing to the coupling model (II), there is an advantage of this coupling model. This coupling model provides a tiny effect of the dynamical evolution in the past since $s \rightarrow 0$. This provides the wider range of the coupling parameter and leads to the very slight deviation of the evolution from the standard evolution of the universe in the past as shown in Figure \ref{evoint3_BP3}.

For coupling (IV), $Q = \sqrt{6}\Gamma (\dot{X} + 3HX)/\kappa$, all of the fixed points A are not stable. Their eigenvalues can be reduced to the non-interacting case. However, we do not explicitly show the eigenvalues because of their complication. For the fixed points B, they can give the stable fixed points at which $\Omega_{X}/\Omega_{c} \sim 7/3$ as shown in Figure \ref{evoint4Bp3}. This provides the ability to solve the coincidence problem while the evolution is in agreement with the standard evolution of the universe.

Even though there is no theoretical motivation for this coupling form, it may be interpreted as a physical interaction associated with particles since it does not depend on the expansion rate of the universe. $\Gamma$ may also be interpreted as a decay rate like $\Gamma$ in coupling model (III). This is one of the advantages for this coupling model.

There is the cosmological constant fixed point, which $\lambda = 0$, for coupling model (I), (II), (III) and non-interacting case but it does not exist for coupling model (IV). The important properties of this fixed point are that it is the stable fixed point with $\rho_X = \text{constant}$ and $\omega_X = \text{constant} = -1$. This is the property of cosmological constant dark energy and leads to the result that there are no stable fixed points at which $\Omega_{X}/\Omega_{c} \sim O(1)$. Covering the cosmological constant dark energy by setting $\lambda = 0$ or $V = V_0$ is one of the important features for the three-form dark energy model which is distinguished from scalar field dark energy.

By comparing the results to the interacting scalar field with type (II) and (III) \cite{int2 negativeE, int23 negativeE}, it gives the similar result in which the coincidence problem will be solved. However, there is no negative energy density of the dark energy in our coupling model. We also note that there is only one parameter $\gamma$ to adjust in our coupling model while it has to have two parameters, $\alpha_\phi$ and $\alpha_c$ in the scalar interacting case. This suggests that it is easier to provide the possibility to solve the coincidence problem in interacting three-form field. One of the advantages of our model is that it automatically provides $\omega_{\text{tot}}$ slightly less than $-1$ before it reaches the stable fixed points as shown in Figure \ref{evoint4Bp3}. This leads to the convenient way to fit the result with the observational data since recent observations suggest that $\omega = -1.1 \pm 0.14$ \cite{wmap7}.

\section*{Acknowledgments}
\indent We would like to thank Ahpisit Ungkitchanukit and Khamphee Karwan for valuable discussions. We are especially grateful to Auttakit Chatrabhuti for his guidance and insightful discussions.

\end{document}